\documentclass{aa}  

\usepackage{graphicx,nicefrac}
\usepackage{txfonts}
\usepackage{hyperref}
\usepackage{color}
\usepackage{pdflscape}
\usepackage{multirow}
\usepackage[capitalise]{cleveref}

\DeclareRobustCommand{\minsize}{\texttt{min\_cluster\_size}}
\DeclareRobustCommand{\minsamp}{\texttt{min\_samples}}
\DeclareRobustCommand{\eps}{\texttt{cluster\_selection\_epsilon}}

\DeclareRobustCommand{\ispec}{\texttt{iSpec} }

\DeclareRobustCommand{\rgc}{R_{\mathrm{GC}}}

\begin{document}

   \title{The (im)possibility of strong chemical tagging}

   \author{L. Casamiquela\inst{1}
          \and A. Castro-Ginard\inst{2,3}
          \and F. Anders\inst{2}
          \and C. Soubiran\inst{1}}

   \institute{Laboratoire d’Astrophysique de Bordeaux, Univ. Bordeaux, CNRS, B18N, allée Geoffroy Saint-Hilaire, 33615 Pessac, France
    \email{laia.casamiquela-floriach@u-bordeaux.fr}
   \and 
   Dept. Física Quàntica i Astrofísica, Institut de Ciències del Cosmos (ICCUB), Universitat de Barcelona (IEEC-UB), Martí Franquès 1, E08028 Barcelona, Spain
   \and 
   Leiden Observatory, Leiden University, Niels Bohrweg 2, 2333 CA Leiden, Netherlands}{}

   \date{Received ; accepted }

  \abstract
   {The possibility of identifying co-natal stars that have dispersed into the Galactic disc based on chemistry only is called strong chemical tagging. Its feasibility has been debated for a long time, with the promise of reconstructing the detailed star-formation history of a large fraction of stars in the Galactic disc.}
   {We investigate the feasibility of strong chemical tagging using known member stars of open clusters.}
   {We analysed the largest sample of cluster members that have been homogeneously characterised with high-resolution differential abundances for 16 different elements. We also investigated the possibility of finding the known clusters in the APOGEE DR16 red clump sample with 18 chemical species. For both purposes, we used a clustering algorithm and an unsupervised dimensionality reduction technique to blindly search for groups of stars in chemical space.}
   {Even if the internal coherence of the stellar abundances in the same cluster is high, typically 0.03 dex, the overlap in the chemical signatures of the clusters is large. In the sample with the highest precision and no field stars, we only recover 9 out of the 31 analysed clusters at a 40\% threshold of homogeneity and precision. This ratio slightly increases when we only use clusters with 7 or more members. In the APOGEE sample, field stars are present along with four populated clusters. In this case, only one of the open clusters was moderately recovered.}
   {In our best-case scenario, more than 70\% of the groups of stars are in fact statistical groups that contain stars belonging to different real clusters. This indicates that the chances of recovering the majority of birth clusters dissolved in the field are slim, even with the most advanced clustering techniques. We show that different stellar birth sites can have overlapping chemical signatures, even when high-resolution abundances of many different nucleosynthesis channels are used. This is substantial evidence against the possibility of strong chemical tagging. However, we can hope to recover some particular birth clusters that stand out at the edges of the chemical distribution.}


   \maketitle
%
\section{Introduction}

The chemical tagging technique \citep{Freeman+2002} consists of grouping stars with similar chemical signatures. \emph{Weak} chemical tagging has been used to find stars that were born in similar Galactic environments and that therefore show similar chemical patterns. This is the case for the identification of large structures such as the thick disc \citep{Hawkins+2015} and also for more particular stellar structures such as the N-rich population in the inner Galaxy \citep{Schiavon+2017}. In contrast, the idea behind \emph{strong} chemical tagging is to find stars that were born in the same star-forming event, that is, in the same birth cluster.

There is a consensus in the community that most of the stars that we see today in the disc were born in stellar aggregates (i.e. open clusters or unbound associations). Star formation simulations and observations tell us that in the process of forming stars, parent molecular clouds undergo fragmentation, thus producing hundreds of stars in the same burst \citep[e.g.][]{Krumholz+2014}. As the result of a number of dynamical processes, most of these aggregates later tend to disperse into the disc in a few $\sim100$ Myr \citep{Krumholz+2019}. However, the highly dissipative nature of the dynamical interactions in the disc prevents us from using the observed kinematics of the individual stars to track them back to their common formation sites. Nevertheless, stars preserve their birth chemical information in their stellar atmospheres for most chemical elements. Assuming a uniform composition of the parent molecular cloud, we can therefore hope to associate individual stars with their birth clusters using chemistry alone. 
This is the idea behind strong chemical tagging, and it has been one of the motivations of several spectroscopic surveys, including APOGEE \citep{Majewski+2017}, GALAH \citep{DeSilva+2015} or the \emph{Gaia}-ESO survey \citep{Randich+2013, Gilmore+2012}. Known open clusters (OCs) are the perfect testbed for studying the possibilities of strong chemical tagging because they are the only example of birth clusters that have survived dynamical effects and remain gravitationally bound today.

Two assumptions need to be confirmed in order to enable strong chemical tagging: i) the members of a birth cluster should have a chemically homogeneous composition, and ii) each cluster should have a unique chemical signature to be able to distinguish stars from different clusters.

The level of chemical homogeneity in OCs has been studied in recent years. It is known that in some cases, cluster members at different stages of stellar evolution can present differences in their abundances. This is for example the case of turnoff stars \citep[e.g.][]{BertelliMotta+2018,Souto+2019,Liu+2019}, where inhomogeneities as high as 0.1-0.2 dex are attributed to diffusion. Additionally, the process of planet formation can result in variations in surface chemical abundances of the host star \citep{Melendez+2009b, Liu+2016b, Spina+2018b}. This effect is usually small and has only been detected using high-precision abundances of solar twins. Most of the studies indicate that OCs have a uniform chemical composition of FGK main-sequence or red giant stars, at least up to $\sim0.02-0.03$ dex \citep{DeSilva+2006, Liu+2016, Bovy2016, Casamiquela+2020}, which is below the level of the uncertainties of large spectroscopic surveys. 

The second requirement for the viability of strong chemical tagging is yet to be proved.
Some studies tried to chemically tag stars in known OCs blingly using high-precision abundances \citep{Mitschang+2013, BlancoCuaresma+2015}, finding it difficult to identify co-natal groups of stars through automated algorithms. \citet{Smiljanic+2018} applied a hierarchical clustering algorithm to eight OCs with FGK type stars in the Gaia-ESO survey data, finding some chemical separation in only five out of eight clusters. \citet{Kos+2018} used the GALAH abundances and the algorithm t-distributed stochastic neighbour embedding (t-SNE) to visually identify nine known open and globular clusters within the field stars. They concluded that t-SNE isolates the different clusters in their chemical space well, even though they did not attempt to run any clustering algorithm. \citet{GarciaDias+2019} used APOGEE DR14 data to test the capability of several non-hierarchical clustering algorithms to distinguish 23 known open and globular clusters with at least five members in different evolutionary states. Their best results without constraining the number of clusters used DBSCAN\footnote{Density-based spatial clustering of applications with noise \citep{Ester+1996}} , which makes a homogeneity\footnote{The homogeneity score measures at which level the predicted clusters contain only data points that are members of one real cluster. A value of 1 means that the clusters are perfectly homogeneous.} score of 0.853. They argued that the primary source of confusion are clusters with similar ages. This means that the algorithm can probably separate open from globular clusters well, but cannot distinguish the different OCs of similar age. They concluded that with the chemical information provided by APOGEE \citep[abundances obtained from the H-band spectra at a spectral resolution of 22,000][]{GarciaPerez+2016}, it is not possible to completely distinguish all the stellar clusters from each other.
After these results, it might be wondered (1) what would happen in the even more ideal case when higher resolution spectra of cluster stars in exactly the same evolutionary stage were used (to avoid systematic biases in the abundances). Could we distinguish cluster stars in this case? 
Aleternatively, (2) what would happen if one includes field stars to the known open cluster sample? this case would possibly increase the difficulty of the experiment but would resemble more the exercise of finding dissolved clusters in the field.

Linking the last idea, a recent study by \citet{PriceJones+2020} claimed to have found several candidate star clusters that were dissolved in the Galactic disc. They appied the DBSCAN algorithm to the APOGEE DR16 sample. However, it is not clear from this study whether their technique can recover the clusters \citep[see][]{Donor+2020} that are known to be present in APOGEE. At this point, it is therfore mandatory to agree on the viability or impossibility of strong chemical tagging to decide whether the candidate star clusters that were found are reliably detected, which applies also to those that may be found in future spectroscopic surveys.

The present work aims to investigate the viability of strong chemical tagging using OCs.
First, we present the best-case scenario using a sample of 31 clusters with at least 4 stars in the same evolutionary stage characterised with strictly line-by-line differential abundances. This represents the idea of point (1) mentioned in the previous paragraph.
We then blindly test whether we can differentiate the known clusters present in the APOGEE DR16 release with more than 4 observed stars, following the idea (2). We use the red clump star sample defined by \citet{Bovy+2014}, which includes field stars in the red clump evolutionary stage (RC).

We organise the paper as follows. In Sect.~\ref{sec:data} we give the details of the abundance data we used and the quality cuts we made. In Sect.\ref{sec:method} we explain the clustering method that we used. Sect.~\ref{sec:EVOC} contains the results of the tests we made for case (1), the high-precision sample, and Sect.~\ref{sec:APOGEE} reports the results for the APOGEE DR16 sample. Finally, we discuss the implications of the results in Sect.~\ref{sec:discussion}, and the main conclusions of the paper are summarised in Sect.~\ref{sec:conclusions}.

\section{Data}\label{sec:data}
For the first part of our analysis (Sect.~\ref{sec:EVOC}), we used the high-precision abundance data published in \citet{Casamiquela+2021} for RC stars belonging to 47 OCs. In this study, the authors analysed high-resolution spectra ($>45,000$) of high-probability members in the RC phase belonging to different clusters. They obtained 1D local thermodynamic equilibrium (LTE) abundances of 25 different chemical species using spectral synthesis fitting with an adapted pipeline that runs the public spectroscopic software \ispec \citep{BlancoCuaresma+2014, BlancoCuaresma2019}. We refer to the original paper for more details concerning the analysis. The fact that the analysed stars correspond to the same evolutionary stage combined with the strictly line-by-line differential analysis allows us to erase most systematic effects in the usual abundance computations (e.g. blends, non-LTE effects, and poor atomic characterisation).

\begin{figure*}
\centering
\includegraphics[width=0.49\textwidth]{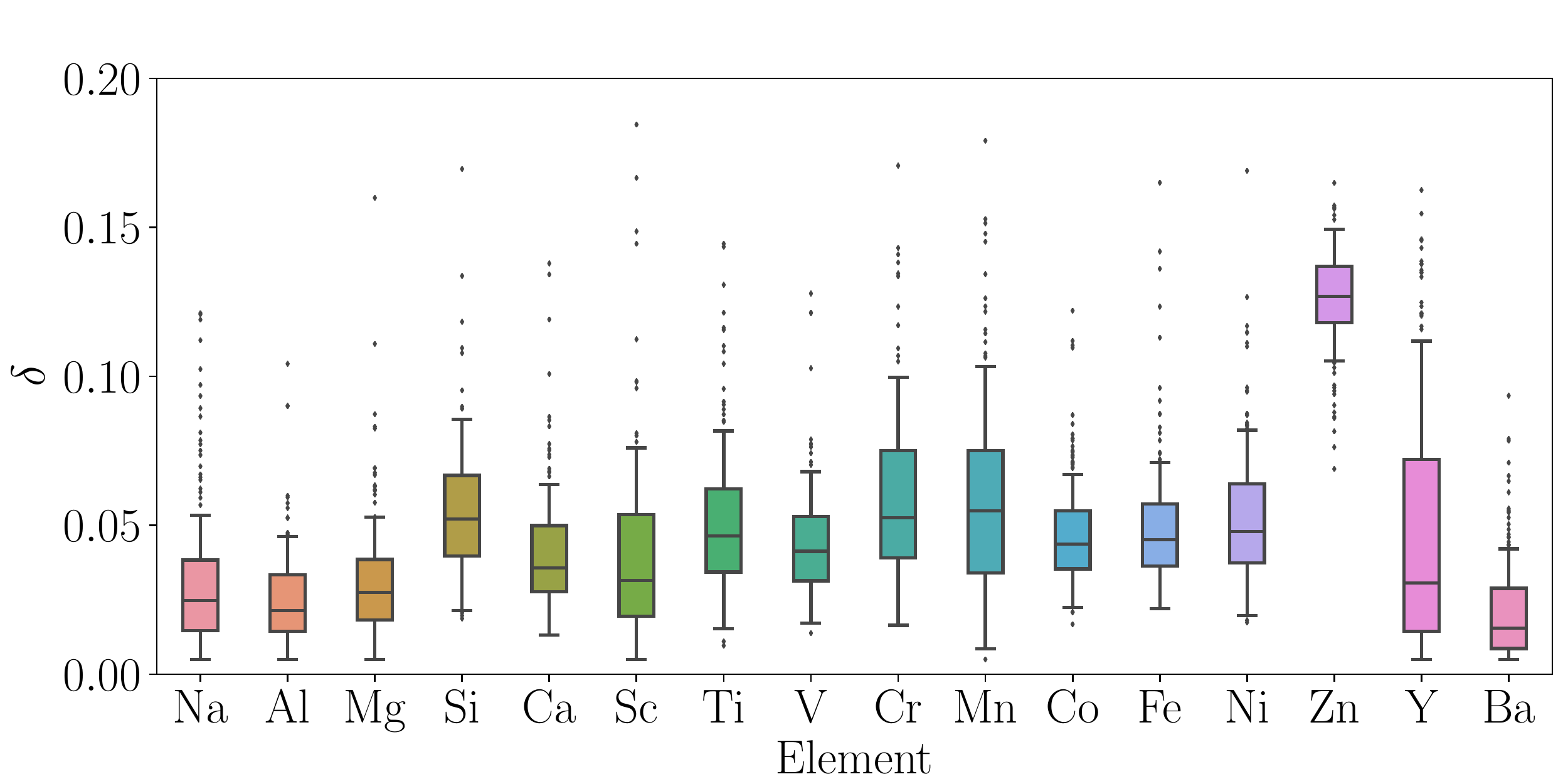}
\includegraphics[width=0.49\textwidth]{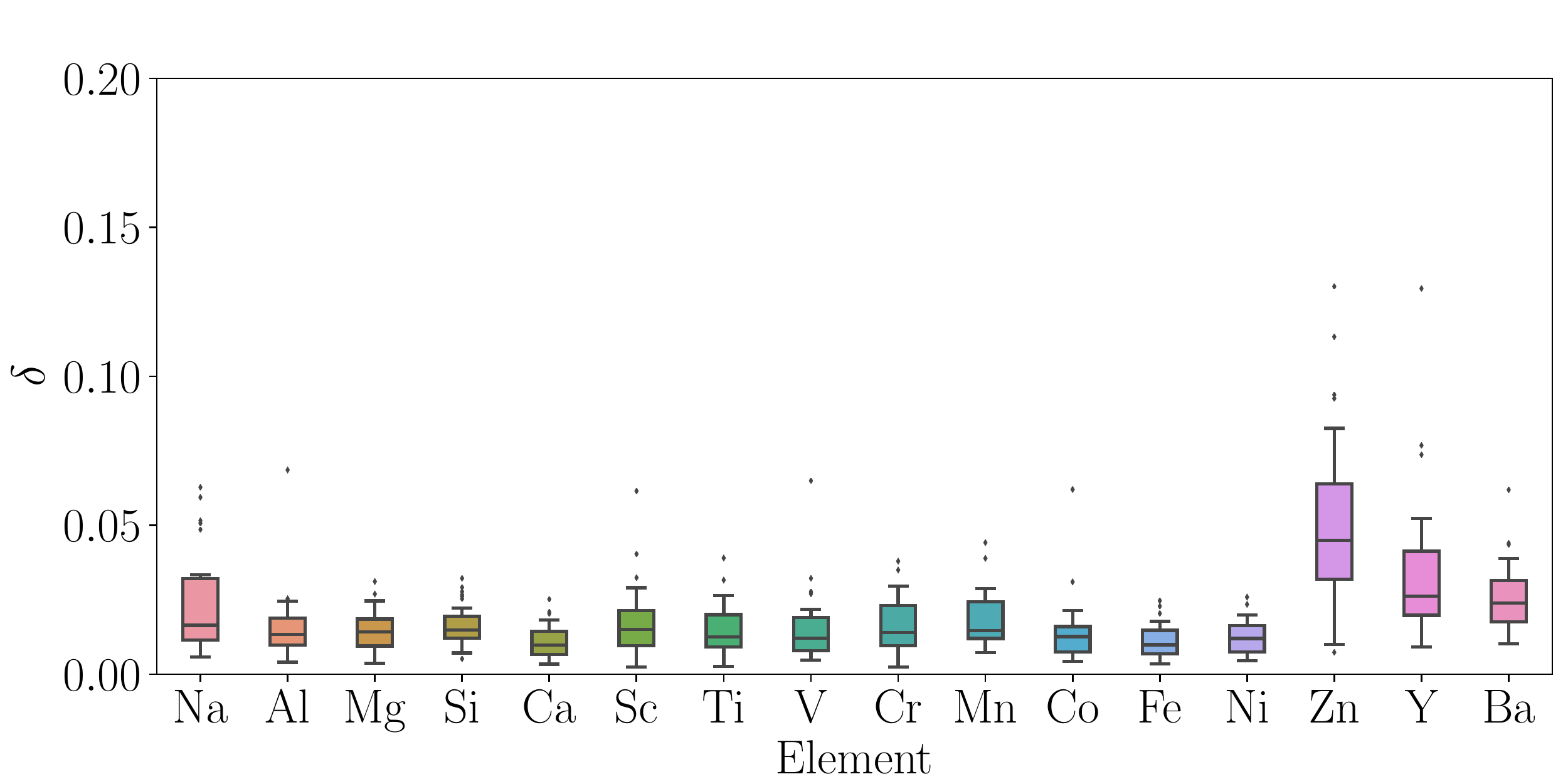}
\caption{Distribution of the abundance uncertainties of the different elements for the high-precision sample \citep{Casamiquela+2021}. Left: Abundance uncertainties of individual stars. Right: Cluster uncertainties taken to be the weighted standard deviation of the abundances for the cluster members.}\label{fig:uncert}
\end{figure*}

For the present work, we used the subsample of clusters that have 4 or more observed stars with complete chemical information, considering 16 chemical species (which is a compromise to maximise the number of stars). The restricted chemical space consists of elements coming from different nucleosynthetic paths: Na, Al, Mg, Si, Ca, Sc, Ti, V, Cr, Mn, Co, Fe, Ni, Zn, Y, and Ba. Even though the results of \citet{Casamiquela+2021} included more chemical species, we selected the elements that have lower uncertainties. Our final sample includes 175 stars in 31 clusters. The individual abundance uncertainties in these elements are usually about 0.05 dex or lower, except for Zn, which is typically between 0.10-0.15 dex (see Fig.~\ref{fig:uncert}). The distribution of the cluster dispersions in abundances is plotted in the right panel of Fig.~\ref{fig:uncert}. It shows that the internal coherence of the stellar abundances in the same cluster is high, typically 0.03 dex. This sample ("high-precision sample", hereafter) provides the best-case scenario for testing chemical tagging: high-precision differential abundances of 16 different elements, and high-probability member stars at the same evolutionary stage.

For the second part of our analysis (Sect.~\ref{sec:APOGEE}), we used APOGEE DR16 \citep{Ahumada+2020} data, which contain detailed abundances from infrared spectra (at a spectral resolution $R\sim22,000$) from over 100,000 stars across the Milky Way computed with the APOGEE Stellar Parameters and Abundances Pipeline \citep[ASPCAP,][]{GarciaPerez+2016}. The details of the data reduction and calibration applied for DR16 data are described in \citet{Jonsson+2020}.
We used the APOGEE abundances obtained with \emph{astroNN} \citep{Leung+2019}, a neural network that was trained on the results of ASPCAP from APOGEE for spectra with a high signal-to-noise ratio (S/N). \emph{astroNN} produces high-precision parameters and abundances for all stars in APOGEE. This was the sample of abundances that \citet{PriceJones+2020} used as well, in which several candidate birth clusters were identified. We used the updated \emph{astroNN} results\footnote{\url{https://www.sdss.org/dr16/data_access/value-added-catalogs/?vac_id=the-astronn-catalog-of-abundances,-distances,-and-ages-for-apogee-dr16-stars}}, trained on APOGEE DR16 abundances.

The sample used in this study is the APOGEE DR16 RC catalogue \citep{Bovy+2014, Ahumada+2020}. This sample contains over 39,000 targets and was built by imposing several criteria on the computed stellar atmospheric parameters, metallicities, and photometry.  The sample contains abundances of 26 chemical species. We have applied quality cuts in the data: S/N$>100$, $\chi^2$ of the fit $<25$, and additional cuts to avoid telluric objects, emission stars, etc\footnote{ASPCAPFLAGS = [BAD, NO\_ASPCAP]; TARGFLAGS = [TELLURIC, SERENDIPITOUS, MASSIVE, EMISSION]; and STARFLAGS = [BAD, COMMISSIONING, SUSPECT]}.
We additionally selected the more reliable elements for the red giant stars: C, N, O, Na, Mg, Al, Si, P, S, K, Ca, Ti, V, Cr, Mn, Fe, Co, and Ni. Additionally, we used the element-wise flag "ELEMFLAG" to avoid problematic abundances\footnote{\url{https://www.sdss.org/dr16/irspec/abundances/}}. Hereafter, we refer to this cleaned sample as the APOGEE DR16 RC sample.

\begin{figure}
\centering
\includegraphics[width=0.49\textwidth]{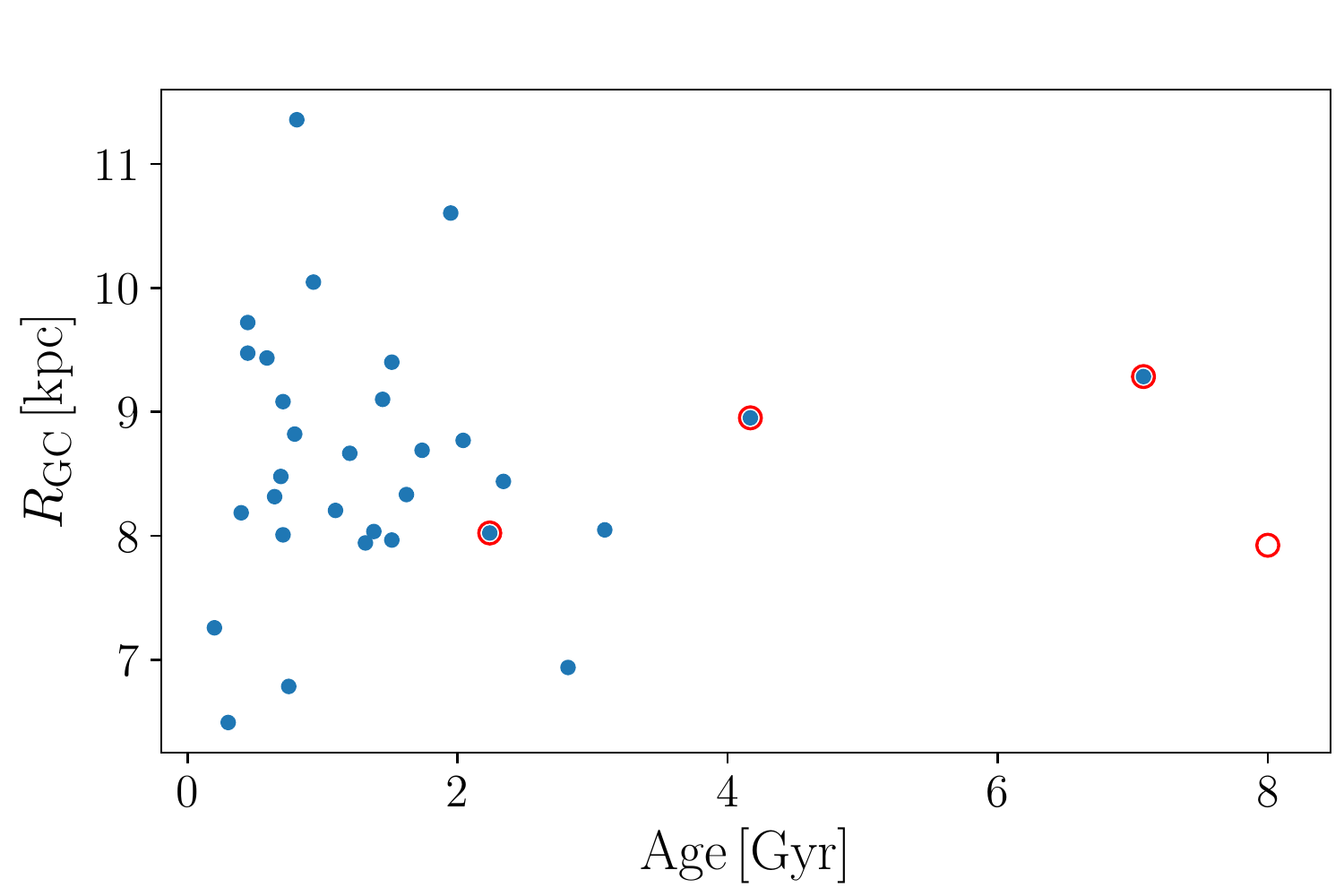}
\caption{$\rgc$-age distribution of the samples of clusters. The high-precision clusters are shown in blue, and the clusters in the APOGEE DR16 RC catalogue are plotted in red.}\label{fig:rgc_age-dist}
\end{figure}

We show in Fig.~\ref{fig:rgc_age-dist} the $\rgc$-age distribution of the two samples of clusters. They span a range of galactocentric distance of $\sim7-11$ kpc, and they have ages mainly younger than 4 Gyr. Only 3 clusters are older than this. Three clusters are in common in the two samples.

\section{Method: HDBSCAN}\label{sec:method}
To group stars and find different clusters, we used a density-based clustering algorithm that blindly searches for stellar overdensities in the chemical space. We used a hierarchical version of DBSCAN, which was previously used to identify several hundreds of new OCs in the {\em Gaia} astrometric parameter space \citep{Castro-Ginard+2018, Castro-Ginard+2019, Castro-Ginard+2020}. DBSCAN relies on two hyper-parameters, $\epsilon$ and \texttt{min\_cluster\_size}, to define a density threshold and detect clusters with a density higher than this threshold. In practice, it defines an $\epsilon$-neighbourhood around each star and searches for other stars within this neighbourhood. If a sufficient number of stars, defined by \texttt{min\_cluster\_size}, fall in the $\epsilon$-neighbourhood, this group is considered as a cluster. The group grows by repeating the process for all stars that can be reached \citep[see details in][]{Castro-Ginard+2018}. With these properties, DBSCAN does not require an \textit{\textup{a priori}} number of clusters to find, it can find clusters with arbitrary shapes, and it can deal with stars that  are not associated with any cluster (noise). The main disadvantage is that DBSCAN is limited to a single density threshold, which usually corresponds to the densest cluster in the field. The hierarchical DBSCAN algorithm \citep[HDBSCAN,][]{Campello+2013} takes advantage of the DBSCAN method, but applies it over all the different $\epsilon$ possibilities. Therefore HDBSCAN is able to find clusters with varying densities, which solves the main disadvantage of DBSCAN while retaining the remaining advantages. 

We used the implementation of HDBSCAN in the python library of the same name\footnote{\url{https://github.com/scikit-learn-contrib/hdbscan}}. In this implementation, three parameters can be fine-tuned in the algorithm, which affects the clustering. The main parameter, \minsize,\, is the most intuitive and refers to the smallest size grouping that we wish to consider as a cluster, as in the aforementioned case of DBSCAN. The other two parameters are due to the choice of the implementation. The \minsamp\ parameter sets how conservative we wish the clustering to be: a low value will make the algorithm sensitive to more local density fluctuations, while a high value captures a more global picture. This parameter is highly data dependent, but it is usually set by default as equal to the \minsize\ parameter. The last parameter that we considered is the \eps.\, Setting a value for it ensures that clusters below a given density threshold defined as the number of stars in an $\epsilon$-neighbourhood are not split any further. This value depends on the given n-dimensional distances among the different data points.

The choice of the free parameters of HDBSCAN is highly dependent on the data that are used and on the purpose. We discuss our choice of the free parameters of HDBSCAN according to our data in Sect.~\ref{sec:chemtag}.

\section{High-precision sample}\label{sec:EVOC}
Well-known OCs are the perfect test case for verifying chemical tagging algorithms with the purpose of finding co-natal stars. In this section, we use the differential chemical abundances obtained by \citet{Casamiquela+2021}, described in the Sect.~\ref{sec:data} as the high-precision sample. In Table~\ref{tab:ocs} we list the 31 clusters with their properties as listed in \citet{Casamiquela+2021}, and the number of member stars in the RC.

\begin{table}
\caption{Selection of clusters, taken from the differential abundance analysis of \citet{Casamiquela+2021}. We indicate the age, the mean cluster metallicity, and the number of RC stars in each cluster.}\label{tab:ocs}
\centering
\setlength\tabcolsep{2pt}
\begin{tabular}{lccc}
 \hline
Cluster  &  NumStars & Age [Gyr] & [Fe/H] \\
\hline
UBC 3       &  4  &0.20&  $-0.081\pm  0.006$ \\
NGC 6705      &  12 &0.30&  $0.047 \pm 0.004$ \\
NGC 3532      &  5  &0.39&  $-0.034\pm  0.012$ \\
UBC 215       &  5  &0.44&  $-0.005\pm  0.023$ \\
NGC 2099      &  10 &0.44&  $-0.003\pm  0.009$ \\
NGC 7245      &  4  &0.58&  $-0.002\pm  0.013$ \\
NGC 6728      &  5  &0.75&  $-0.065\pm  0.020$ \\
NGC 6997      &  6  &0.64&  $0.096 \pm 0.008$ \\
NGC 2632      &  4  &0.69&  $0.118 \pm 0.014$ \\
NGC 6633      &  4  &0.70&  $-0.043\pm  0.005$ \\
NGC 2539      &  5  &0.70&  $-0.012\pm  0.004$ \\
UBC 6       &  6  &0.79&  $-0.030\pm  0.005$ \\
NGC 2266      &  5  &0.81&  $-0.057\pm  0.025$ \\
NGC 2355      &  6  &0.93&  $-0.138\pm  0.010$ \\
NGC 6811      &  6  &1.09&  $-0.059\pm  0.007$ \\
NGC 7245      &  7  &1.20&  $-0.053\pm  0.006$ \\
IC 4756       &  9  &1.31&  $-0.063\pm  0.006$ \\
NGC 6940      &  6  &1.38&  $0.009 \pm 0.010$ \\
NGC 2354      &  6  &1.44&  $-0.144\pm  0.018$ \\
Skiff J1942+38.6  &  4  &1.51&  $-0.002\pm  0.009$ \\
NGC 7789      &  6  &1.51&  $-0.047\pm  0.018$ \\
NGC 6991      &  4  &1.62&  $-0.068\pm  0.009$ \\
NGC 6939      &  5  &1.73&  $-0.032\pm  0.015$ \\
NGC 2420      &  7  &1.95&  $-0.186\pm  0.017$ \\
NGC 7762      &  5  &2.04&  $-0.051\pm  0.010$ \\
NGC 6819      &  4  &2.23&  $-0.050\pm  0.010$ \\
FSR 0278      &  5  &2.34&  $0.024 \pm 0.007$ \\
Ruprecht 171    &  6  &2.81&  $-0.041\pm  0.014$ \\
Ruprecht 147    &  4  &3.09&  $0.053 \pm 0.015$ \\
NGC 2682      &  6  &4.16&  $-0.075\pm  0.007$ \\
NGC 188       &  4  &7.07&  $-0.030\pm  0.015$ \\

\hline
\end{tabular}
\end{table}

\begin{figure*}
\centering
\includegraphics[width=\textwidth]{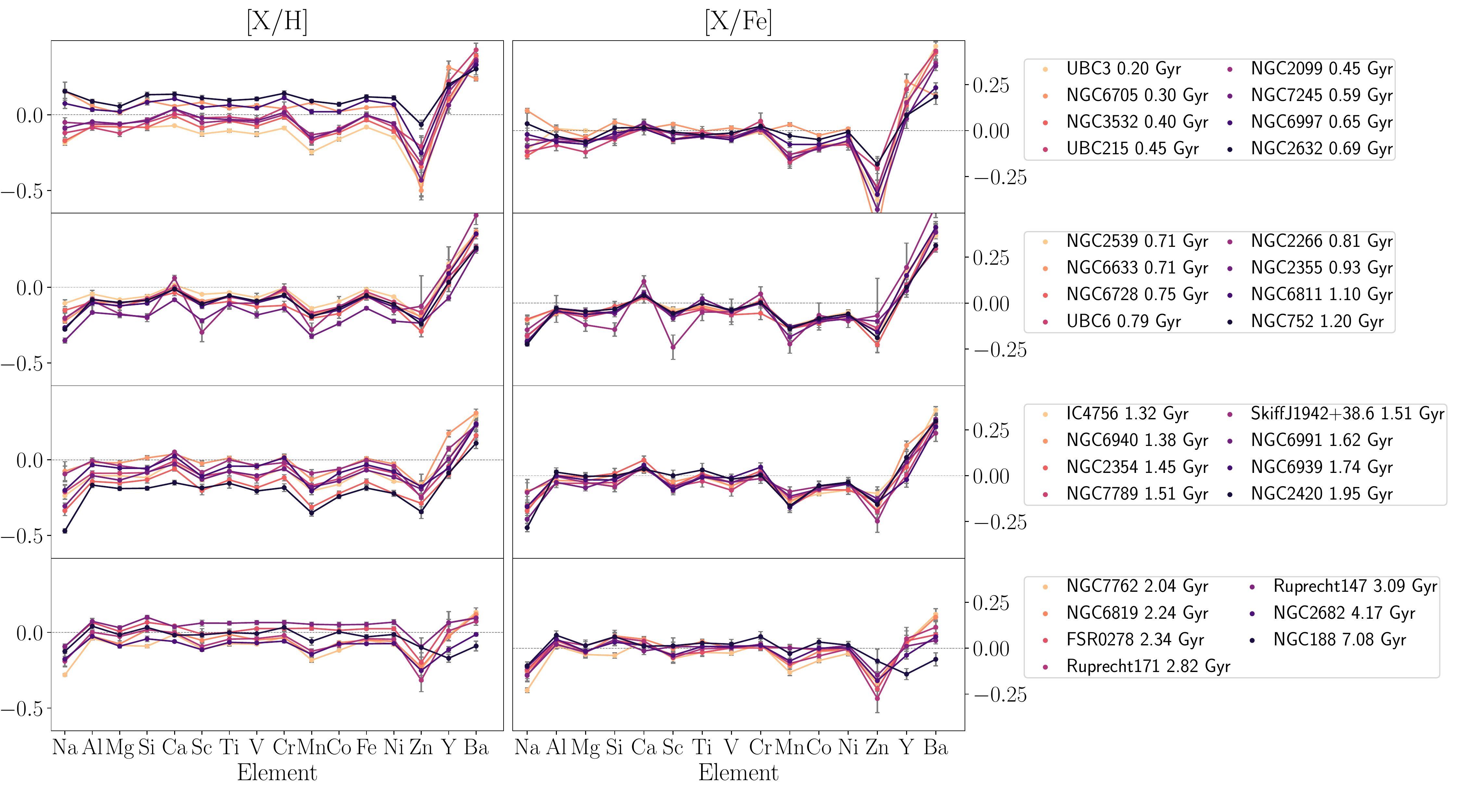}
\caption{Mean [X/H] (left) and [X/Fe] (right) abundances of the sample of 31 clusters in the high-precision sample. Clusters are sorted by age (increasing downwards) in the different panels. They are coloured by age from young (yellow) to old (black).}\label{fig:allabus}
\end{figure*}

In Fig.~\ref{fig:allabus} we plot the cluster abundances [X/H] and [X/Fe] for the full set of clusters. The cluster abundances were computed using a mean weighted by star uncertainties, and the error is the weighted standard deviation of the member stars. The clusters are sorted by age in the different panels. The irregular distribution in age among the different panels complicates a visual interpretation, but in a general picture, the analysed OCs tend to have a nearly solar abundance pattern, except for some chemical species, highlighting Zn, Y, and Ba. This is the case for the youngest clusters, which show a remarkable depletion in Zn abundance and an enhancement in the s-process elements Y and Ba, which returns to solar values with increasing age. The s-process abundance enhancement, particularly that of Ba, is a known effect that has been analysed before \citep[e.g.][]{DOrazi+2009, Maiorca+2012, Magrini+2018, Casamiquela+2021}. It is interesting to include these elements in our abundance values because due to their special behaviour with age, they might eventually help to distinguish between clusters.

The abundance spread in each cluster is small, typically below 0.03 dex (see the right panel of Fig.~\ref{fig:uncert}), thus the uncertainties in Fig.~\ref{fig:allabus} are usually smaller than the points. Visually, we find that the chemical signatures of the analysed clusters overlap widely and span a small range of abundances (-0.2 to 0.1 dex in [X/H]), with a few exceptions. Our sample of clusters has a relatively wide range of age (200 Myr to 7 Gyr), but the highly overlapping signatures of the clusters make it difficult to distinguish one cluster from the other. The analysed population essentially represents the latest instants of the evolution of the Milky Way disc. This indicates that in the latest billion years, the interstellar medium from which these clusters formed was well mixed. Our result is consistent with recent studies in other spiral galaxies, such as \citet{Kreckel+2020}, who obtained a low abundance scatter (0.02-0.03 dex) in several nearby disc galaxies. This implies that the spatial metallicity distribution is highly correlated at scales $<600$ pc.

\subsection{Chemical tagging with HDBSCAN}\label{sec:chemtag}
We have run HDBSCAN in the chemical space of 16 elements described in Sect.\ref{sec:data}. This is a controlled sample, where we know which star belongs to which cluster and with no field stars. Thus, it is the perfect case to evaluate the performance of the clustering algorithm. For the mentioned reasons, and contrary to most situations in a blind clustering search, we can apply different diagnostics which are informative of how well the clusters are recovered. We use the indicators:

\begin{enumerate}
  \item The V-measure \citep{Rosenberg+2007} measures how successfully the criteria of homogeneity and completeness have been satisfied in the recovered groups\footnote{Notation throughout the paper: we use \emph{clusters} when we are referring to real open cluster stars, and we use \emph{groups} to designate the clusters found by the algorithm, which are not necessarily real open clusters}. A clustering result satisfies homogeneity ($h$) if all of its groups contain only data points that are members of a single OC. On the other hand, completeness ($c$) is satisfied if all the data points that are members of a given group are elements of the same cluster. The non-weighted V-measure is defined as: $V = \nicefrac{2\cdot h\cdot c}{h+c}$, it equals 1 for a perfectly complete clustering.  We refer the reader to the original paper for the mathematical definitions of the two quantities. Several other papers have used this indicator also for the same purpose \citep[e.g.][]{BlancoCuaresma+2015, GarciaDias+2019}
  \item Similarly to \citet{PriceJones+2019} we define the Recovery Fraction ($RF$) as the fraction of successfully recovered groups with respect to the initial number of real clusters. We consider a cluster successfully recovered if the group it is assigned to exceeds a given homogeneity and completeness thresholds. We first set a more restrictive threshold of 70\%, which is the same as the one used in \citet{PriceJones+2019}. However, we realized that in most test cases this is too restrictive, and at most, only one cluster is recovered with this threshold. Moreover, some clusters moderately recovered were not taken into consideration as successful groups, this was often the case for clusters with few stars. So we compute also a less restrictive recovery fraction with a completeness and homogeneity threshold at 40\%.
\end{enumerate}

\subsubsection*{Fine-tuning the HDBSCAN parameters}\label{sec:tuning}

The main free parameter in HDBSCAN, \minsize,\, fixes the minimum number of members in each group that are to be considered a real group. We chose to fix this to 2. This is a reasonable value in our case because it allows the recovery of at least half of the stars of the real clusters that have the fewest number of stars (four members).

\begin{figure*}
\centering
\includegraphics[width=\textwidth]{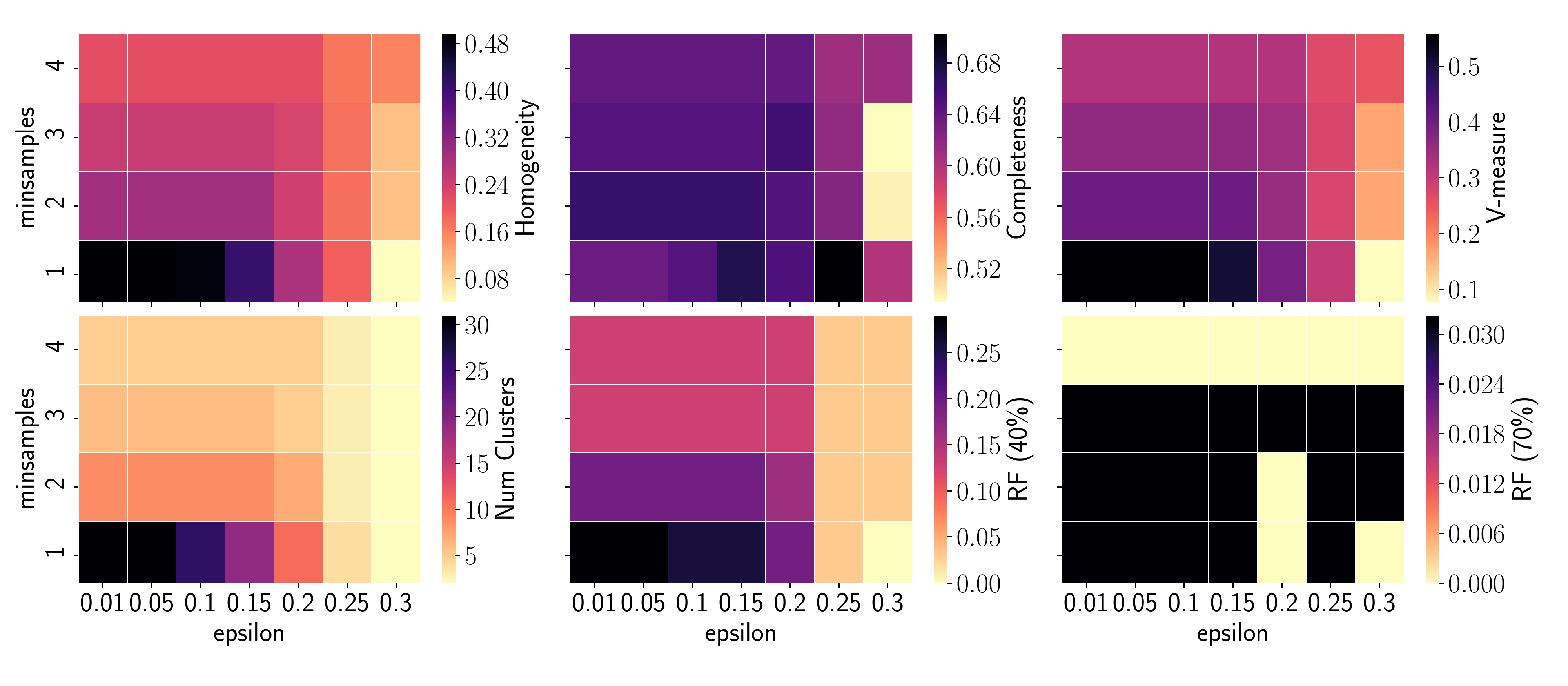}
\caption{Heatmaps showing the results of the quality indicators for a sampling of the HDBSCAN free parameters \minsamp\ and \eps. The top row shows the homogeneity, completeness, and V-measure as defined by \citet{Rosenberg+2007}. In the bottom row, we show the number of found groups and the RF with two different thresholds to the homogeneity and completeness, 70\% and 40\%.}\label{fig:recovery}
\end{figure*}

As explained in Sect.\ref{sec:method}, the performance of the algorithm depends on the two parameters from HDBSCAN that control how conservative the clustering is: \minsamp\ and \eps. To evaluate their effect, we computed the homogeneity, completeness, V-measure, RF (at 40\% and 70\%), and the number of identified groups, sampling $4\times7$ different combinations of these two parameters. The results are shown in Fig.~\ref{fig:recovery} in the form of a heatmap.

The values of the quality indicators show that the clustering appears to be more successful when the \minsamp\ and the \eps\ are small. Setting a low value of these two parameters means that the clustering is less conservative, allowing it to find groups in less dense areas. This results in a larger number of groups that are found, most of them with only two or three stars.
When \eps\ starts to grow ($\geq0.25$), fewer groups are found, and typically one of them contains most of the initial sample of stars, which in reality belong to more than 25 clusters. The same happens for \minsamp\ $\geq2,$ where the algorithm persistently identifies a large group of more than 50 stars belonging to more than 20 real clusters, even if epsilon is small.

In the limit of the highest values of the two parameters, only two or three groups are found. In this case, the algorithm always correctly identifies most of the stars of the cluster NGC 6705 in a single group with 100\% homogeneity and a completeness usually higher than 80\%. The other group typically contains more than 100 stars belonging to more than 30 real clusters. In some cases, however, the group of stars belonging to NGC 6705 is split into two groups when \eps\ $\leq0.2$: both with 100\% homogeneity, but the completeness $\leq50\%$. 
A compromise is needed to fix these two parameters. Lower values allow a larger fragmentation of the clustering space, which enables the recovery of a larger number of real clusters with good homogeneity, but low completeness. In contrast, higher values allow the recovery of the clusters that stand out (NGC 6705 in our case) with high completeness, but all the other few groups that are found typically contain a mixture of stars from different clusters.

For the remainder of the paper, we choose the optimal parameters that maximise the number of correctly recovered clusters, as was done also by \citet{PriceJones+2019}. This will represent the best results we can obtain.
We chose to prioritize the recovery of purer clusters (with a high homogeneity), even if this implies that the recovered completeness will be low. This resembles a blind chemical tagging experiment in a large set of data better, where hundreds of real cluster stars are spread in the field: even if an algorithm can group half of them, this would already allow determining a birth cluster. We therefore chose \minsamp\ $=1$ and \eps\ $=0.05$.
We note that the clustering results are the same for \eps\ $=0.01,\, 0.05$ because this parameter defines the threshold up to which we allow two points to be a group. Because the uncertainties in individual star abundances are about 0.05 (see Fig.~\ref{fig:uncert}), it is more reasonable to choose this last value.

NGC 6705 is recovered persistently in almost all possible configurations. This cluster is known to be particularly enhanced in $\alpha$ elements \citep[e.g.][]{Casamiquela+2018,Magrini+2017}, taking into account its young age ($\sim300$ Myr) and metallicity. It is therefore natural to expect a better recovery of its stars because they are in a region in the chemical space that is detached from the rest of stars. It is also the cluster with the largest number of member stars, which facilitates the recovery.

\subsection{HDBSCAN on the full sample}\label{sec:full_sample}

\begin{figure*}
\centering
\includegraphics[width=\textwidth]{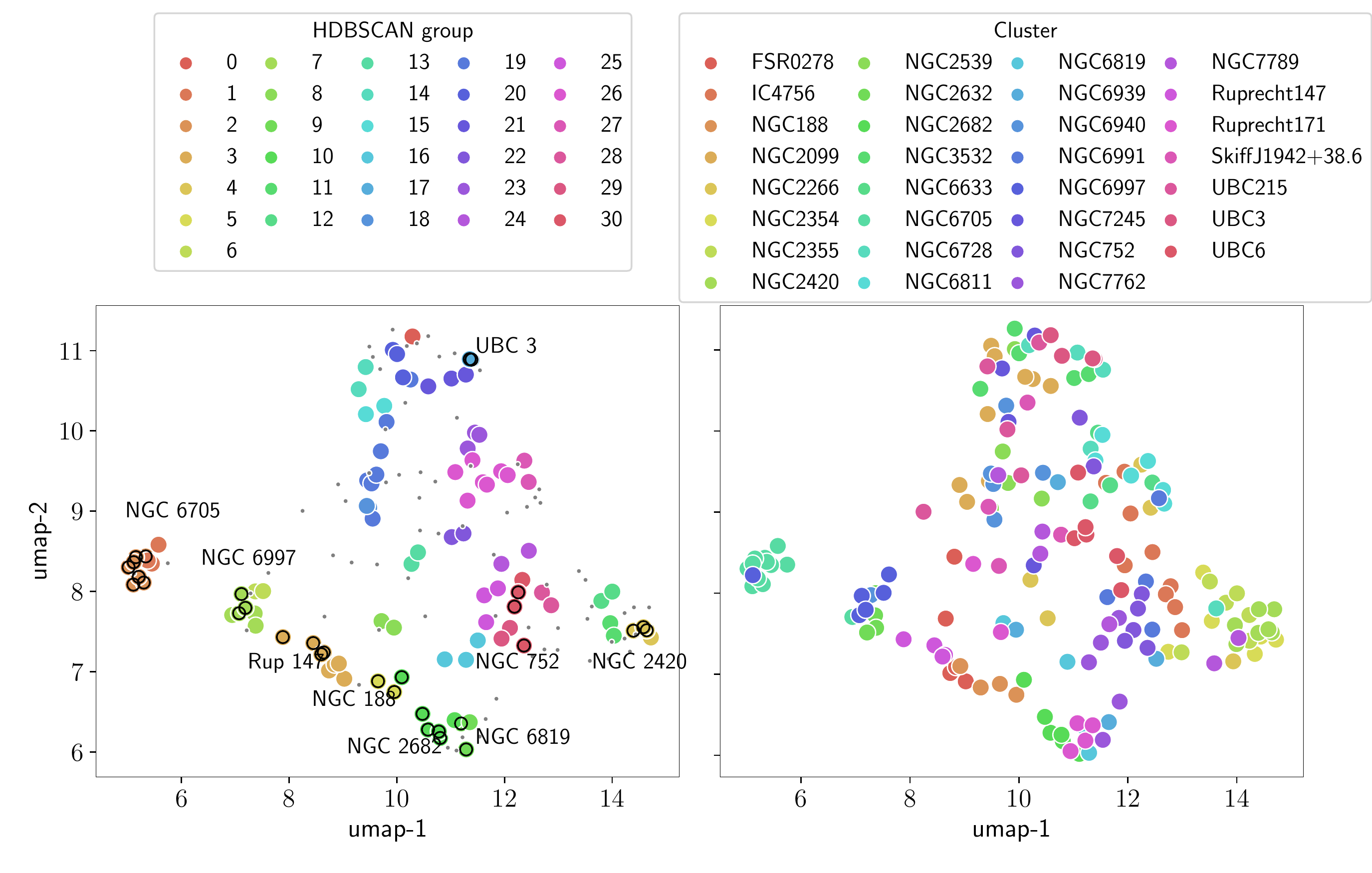}
\caption{UMAP projections of the stars in the high-precision sample. The groups are coloured according to whether they were found by HDBSCAN (left) or are real clusters (right). The grey points correspond to stars that were found as noise. The stars highlighted with black circles were recovered in a group with at least 40\% homogeneity and completeness (those in boldface in Table~\ref{tab:HDBS_clusters}), and the corresponding cluster is labelled.}\label{fig:UMAP}
\end{figure*}

We ran HDBSCAN with the parameters \minsize\ $=2$, \minsamp\ $=1$ and \eps\ $=0.05$. This configuration finds 31 groups with global homogeneity and completeness parameters of 49\% and 63\%, respectively, and a V-measure of 55\%. We find recovery fractions of RF$_{40}=$29\% and RF$_{70}=$3\% for the 40\% and 70\% thresholds, respectively, as explained in Sect~\ref{sec:chemtag}. We show in Table~\ref{tab:HDBS_clusters} the nine groups found by HDBSCAN  with more than 40\% completeness and homogeneity. Out of these, only NGC~2682 is retained when a threshold of 70\% was set. Four of the clusters are found with a homogeneity of 100\%, that is, they were not confused with stars from other clusters, although in most cases, the group contains only two stars of the original cluster. 

In this configuration, the group H2 is selected because it contains seven stars of the cluster NGC~6705. The group H1 contains the other three stars of this cluster, with a 100\% homogeneity but 25\% completeness, but consequently, it is not selected at a threshold of 40\%. As explained in Sect.~\ref{sec:tuning}, these two groups merge in some configurations with higher values of \eps\ or \minsamp. The compromise of lowering the values of the two free parameters to try to recover as many real clusters as possible causes a fragmentation of this cluster.

To better visualise the groups found in the chemical space, we used the algorithm: uniform manifold approximation and projection for dimension reduction \citep[UMAP,][]{McInnes+2018}, which has a library\footnote{\url{https://umap-learn.readthedocs.io/en/latest/}} implemented in python. It has similar objectives as t-SNE or principal component analysis (PCA), which are used in the literature for similar purposes. UMAP performs a reduction of dimensions from 16 to 2, which in our case preserves the global structure of the data to help visualisation and interpretation.

We show the two projections of UMAP for all the stars in our sample in Fig.~\ref{fig:UMAP}. The original clusters are coloured in the right panel, and the groups found by HDBSCAN are coloured in the left panel. The right panel shows that the central part of the distribution has a large mixture of stars from different clusters that are confused, and by eye, it is extremely difficult to distinguish any cluster. However, some structures can already be identified as clumps in this space, for instance, NGC~6705, which stands alone in the leftmost part of the plot.

In the left panel, we highlight with black circles the stars from the groups detailed in Table~\ref{tab:HDBS_clusters}, which are members of a cluster recovered at the 40\% threshold of completeness and homogeneity (the clusters marked in boldface in the table). These stars represent those that we consider \emph{correctly identified}. The plot clearly shows that in most cases, they belong to clusters that are at the edges of the distribution, that is, more separated in chemical space. This includes NGC~6705, but also NGC~2420, the most metal-poor cluster in our sample. In several cases, the stars identified as a group do not appear really clustered in Fig.~\ref{fig:UMAP}, such as NGC~188 and NGC~752, but we point out that in these cases, their completeness (42\% and 50\%, respectively) is close to the limit of our tolerance.

\begin{table}
\caption{Groups found by HDBSCAN in the full sample of clusters in the high-precision sample (see also Fig.~\ref{fig:UMAP}). We only list the groups that represent real clusters at a 40\% threshold in completeness and homogeneity (highlighted in boldface). The number of stars found in each group is indicated, ($N_{\mathrm{FOUND}}$), and for each corresponding real cluster, we also indicate the number of stars found with respect to the total number of cluster stars ($\nicefrac{N_{\mathrm{FOUND}}}{N_{\mathrm{REAL}}}$). For each recovered cluster, the completeness and homogeneity of the recovery is also indicated.}\label{tab:HDBS_clusters}
\centering
\setlength\tabcolsep{2pt}
\begin{tabular}{llrrr}
 \hline
HDBSCAN group  & Real Cluster(s)   & Comp. & Hom.  \\
($N_{\mathrm{FOUND}}$) & ($\nicefrac{N_{\mathrm{FOUND}}}{N_{\mathrm{REAL}}}$) &  & \\
\hline
H2 (7)  & \textbf{NGC 6705} (7/12)   & \textbf{$58\%$} & \textbf{$100\%$} \\
H3 (8)  & \textbf{Ruprecht 147} (4/4)& \textbf{$100\%$} & \textbf{$50\%$} \\
        & NGC 188 (1/4)              & $25\%$ & $12\%$ \\
        & FSR 0278 (3/5)             & $60\%$ & $37\%$ \\
H4 (4)  & \textbf{NGC 2420 (3/7)}    & \textbf{$42\%$} & \textbf{$75\%$} \\
        & NGC 2354 (1/6)             & $16\%$ & $25\%$ \\
H5 (2)  & \textbf{NGC 188 (2/4)}     & \textbf{$50\%$} & \textbf{$100\%$} \\
H7 (6)  & \textbf{NGC 6997} (3/6)    & \textbf{$50\%$} & \textbf{$50\%$} \\
        & NGC 6705 (1/12)            & $8\%$ & $16\%$ \\
        & NGC 2632 (2/4)             & $50\%$ & $33\%$ \\
H9 (4)  & \textbf{NGC 6819} (2/4)    & \textbf{$50\%$} & \textbf{$50\%$} \\
        & NGC 2682 (1/6)             & $16\%$ & $25\%$ \\
        & Ruprecht 171 (1/5)         & $20\%$ & $25\%$ \\
H10 (5) & \textbf{NGC 2682} (5/6)    & \textbf{$83\%$} & \textbf{$100\%$} \\
H17 (2) & \textbf{UBC 3} (2/4)       & \textbf{$50\%$} & \textbf{$100\%$} \\
H30 (4) & \textbf{NGC 752} (3/7)     & \textbf{$42\%$ }& \textbf{$75\%$} \\
        & NGC 6991 (1/4)             & $25\%$ & $25\%$ \\
\hline
\end{tabular}
\end{table}

\subsection{HDBSCAN on clusters with $\geq6$ members}\label{sec:restr_sample}
We realised that some of the clusters that were successfully recovered in the previous test are also those that have a larger number of stars, in particular, NGC 6705. Several clusters in the sample have four or five members, which might statistically be more difficult to recover, and thus this could worsen the recovery fraction.
For this reason, we repeat the same experiment in this subsection with a more restricted case in which only the clusters with at least six members in the high-precision sample are included. This makes a test case with 14 clusters and 99 stars in total.

\begin{figure*}
\centering
\includegraphics[width=\textwidth]{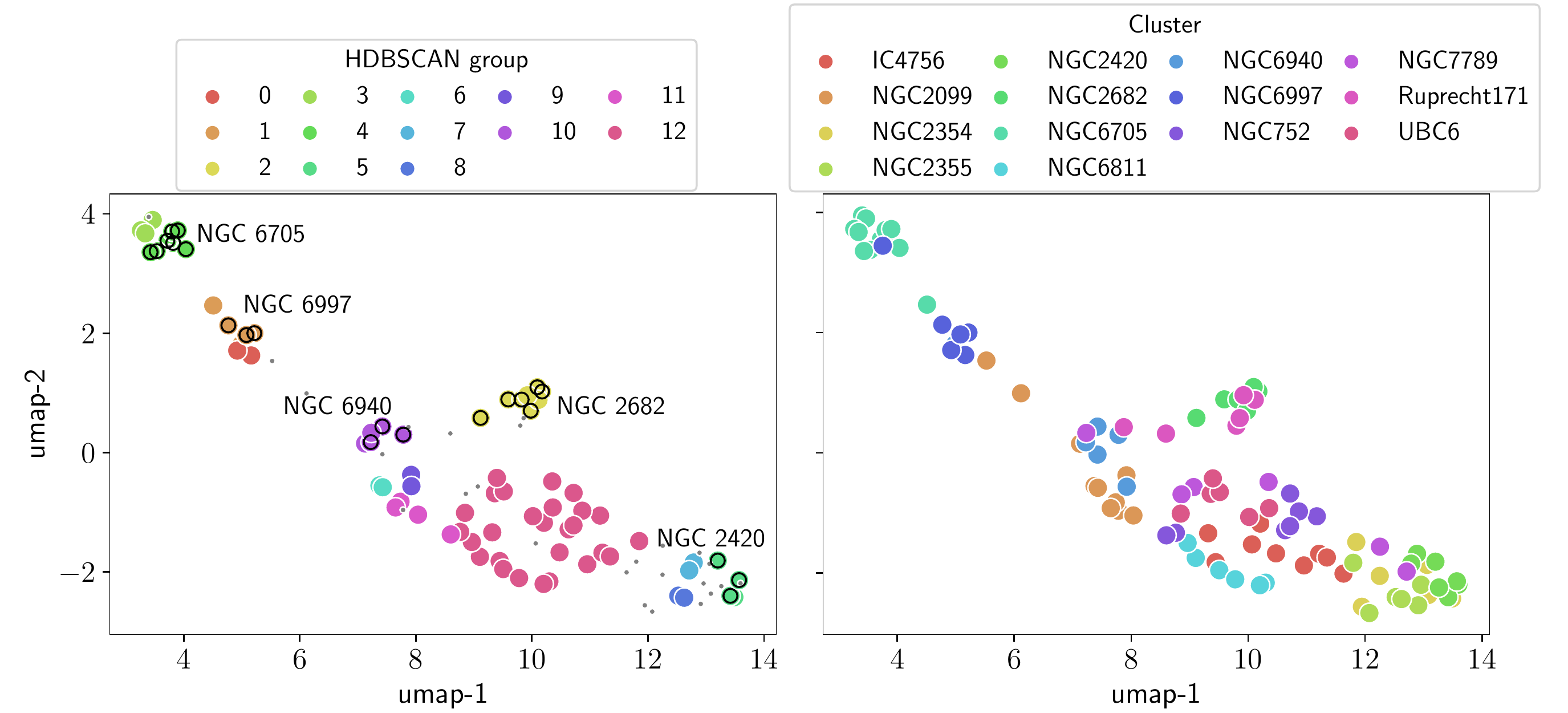}
\caption{Similar to Fig.~\ref{fig:UMAP} but only using clusters with at least 6 members in the high precision sample. UMAP projections of the stars colouring the real clusters (right), and the groups found by HDBSCAN (left). The stars highlighted in black circles are those recovered in a group with at least 40\% homogeneity and completeness (those in boldface in Table~\ref{tab:HDBS_clusters_6m}).}\label{fig:UMAP_6memb}
\end{figure*}

We ran HDBSCAN in the same configuration as in the previous subsection. It found 13 groups in this case, with a global homogeneity and completeness of 53\% and 64\% and a V-measure of 58\%. These values are very similar to those obtained in the previous test: with a homogeneity of 49\%, a completeness of 63\% and a V-measure of 55\%. The recovery fractions in this case are also very similar as before, but slightly larger: RF$_{40}=35$\% and RF$_{70}=7$\% for the 40\% and 70\% thresholds, respectively (to be compared with the RF$_{40}=29$\% and RF$_{70}=3$\% obtained before). We represent the UMAP projections in Fig.~\ref{fig:UMAP_6memb}. The details of the clusters that were successfully recovered at the 40\% threshold (RF$_{40}$) are listed in Table~\ref{tab:HDBS_clusters_6m}.

In this experiment, four out of the five recovered clusters were also recovered when the full high-precision sample was used, and with similar completeness and homogeneity scores. As an additional group, three out of the six stars of NGC 6940 are recovered in the H10, mixed with two stars from two other clusters. Most of the clusters that were recovered in Sect~\ref{sec:full_sample} but that do not appear here have fewer than six members and are not part of the current experiment (Ruprecht 147, NGC 188, NGC 6819, UBC 3).
We wish to highlight that of the three clusters with more stars in the sample (NGC 6705 has 12 stars, NGC 2099 has 10 stars, and IC 4756 has 9 stars), only NGC 6705 is recovered as a successful HDBSCAN group. The stars of the other two are split among the different identified groups, most of them in the H12, which gathers the central region of the UMAP distribution in Fig.~\ref{fig:UMAP_6memb}. Again, the successful groups are in general those that are at the edges of the distribution.

We have further tried to run the algorithm using \minsize\ $=3$ instead of 2, as we set in Sect.~\ref{sec:full_sample}. This would also be a consistent choice to allow the recovery of half of the cluster members with the fewest number stars (which now is 6 instead of 4). This configuration finds 9 groups instead of 13, but the clusters that are successfully recovered are the same as the previous test (Table~\ref{tab:HDBS_clusters_6m}) and with equal homogeneities and completeness.

We also tried to perform the same test using only the clusters with seven or more members. This cut drastically reduces the number of clusters to five, with a total of 44 stars. In this case, at a 40\% threshold on homogeneity and completeness, we recover four clusters (RF$_{40}=80$\%), and at a 70\% threshold, we still only recover one case NGC 6705 (RF$_{70}=20$\%). Again, we recover one group that represents the central part of the distribution, as shown in \cref{fig:UMAP,fig:UMAP_6memb}, with a balanced mixture of stars from IC 4756 and NGC 752. However, we recall that this is a highly unrealistic case of a sample of only a few clusters, which is probably not statistically significant. Even in this case, we can only recover one cluster with at least 70\% on completeness and homogeneity.

\begin{table}
\caption{Same as Table~\ref{tab:HDBS_clusters}, but with the results of the HDBSCAN run on the clusters with six or more members in the high-precision sample  (see also Fig.~\ref{fig:UMAP_6memb}).}\label{tab:HDBS_clusters_6m}
\centering
\setlength\tabcolsep{2pt}
\begin{tabular}{llrrr}
 \hline
HDBSCAN group  & Real Cluster(s)   & Comp. & Hom.  \\
($N_{\mathrm{FOUND}}$) & ($\nicefrac{N_{\mathrm{FOUND}}}{N_{\mathrm{REAL}}}$) &  & \\
 \hline
H1 (5)  & \textbf{NGC 6997 (3/6)}  & \textbf{50\%}  & \textbf{60\%} \\
        & NGC 6940 (1/6)           & 16\%           & 20\% \\
        & NGC 6705 (1/12)          & 8\%            & 20\% \\
H2 (8)  & \textbf{NGC 2682 (6/6)}  & \textbf{100\%}   & \textbf{75\%} \\
        & Ruprecht 171 (2/6)       & 33\%           & 25\% \\
H4 (7)  & \textbf{NGC 6705 (7/12)} & \textbf{58\%}  & \textbf{100\%} \\
H5 (4)  & \textbf{NGC 2420 (3/7)}  & \textbf{42\%}  & \textbf{75\%} \\
        & NGC 2354 (1/6)           & 16\%           & 25\% \\
H10 (5) & \textbf{NGC 6940 (3/6)}  & \textbf{50\%}  & \textbf{60\%} \\
        & NGC 2099 (1/10)          & 10\%           & 20\% \\
        & NGC 7789 (1/6)           & 16\%           & 20\% \\
\hline
\end{tabular}
\end{table}

\section{Finding known clusters in the APOGEE DR16 RC sample}\label{sec:APOGEE}
We now investigate the possibility of finding known clusters in a large dataset for a blind strong chemical tagging experiment with field stars, as designed by \citet{PriceJones+2019}.
We used the APOGEE DR16 RC sample, described in detail in Sect.~\ref{sec:data}, which contains 16,193 stars. This is the most adequate catalogue for our purpose because it contains a large number of stars with good-quality abundances in a very small bin in the HR diagram, so that systematic abundance trends due to the evolutionary state are minimised. This therefore mimics the selection of stars for the clusters in \citet{Casamiquela+2021} that we used in Sect.~\ref{sec:EVOC}.

APOGEE DR16 contains several cluster members. We used the list of observed stars from the 71 high-quality clusters selected by \citet{Donor+2020}, which were selected using Gaia DR2 proper motions and APOGEE radial velocities and [Fe/H]. We cross-matched this table with the APOGEE DR16 RC sample, and we kept only the clusters for which four or more stars were also present in the membership analysis of \citet{Cantat-Gaudin+2018}. This selected only four clusters: NGC 188 (4 stars), NGC 2682 (5 stars), NGC 6791 (6 stars), and NGC 6819 (11 stars). All the selected stars have a probability of membership of at least 0.7 in \citet{Cantat-Gaudin+2018}.

\begin{figure*}
\centering
\includegraphics[width=0.49\textwidth]{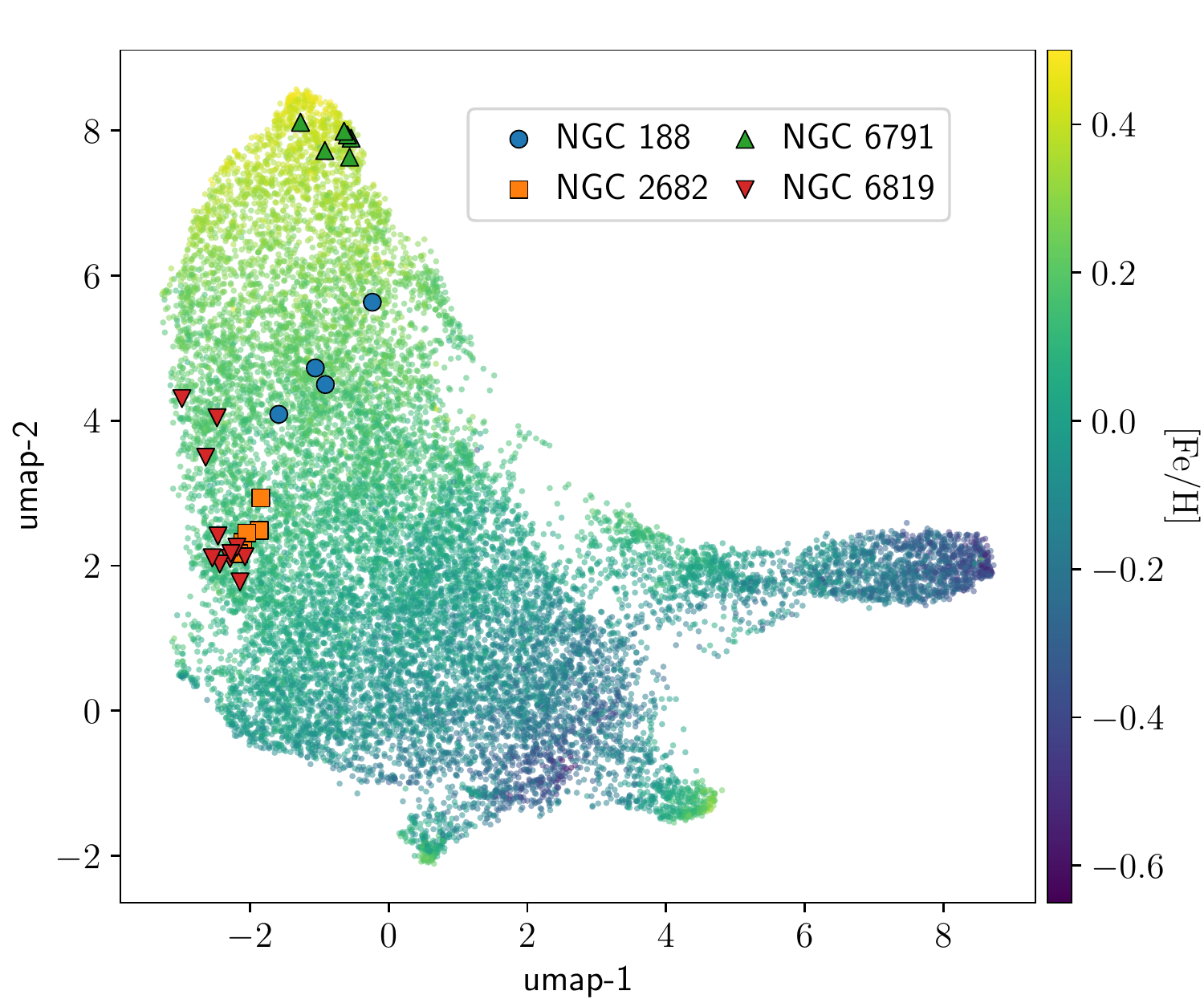}
\includegraphics[width=0.49\textwidth]{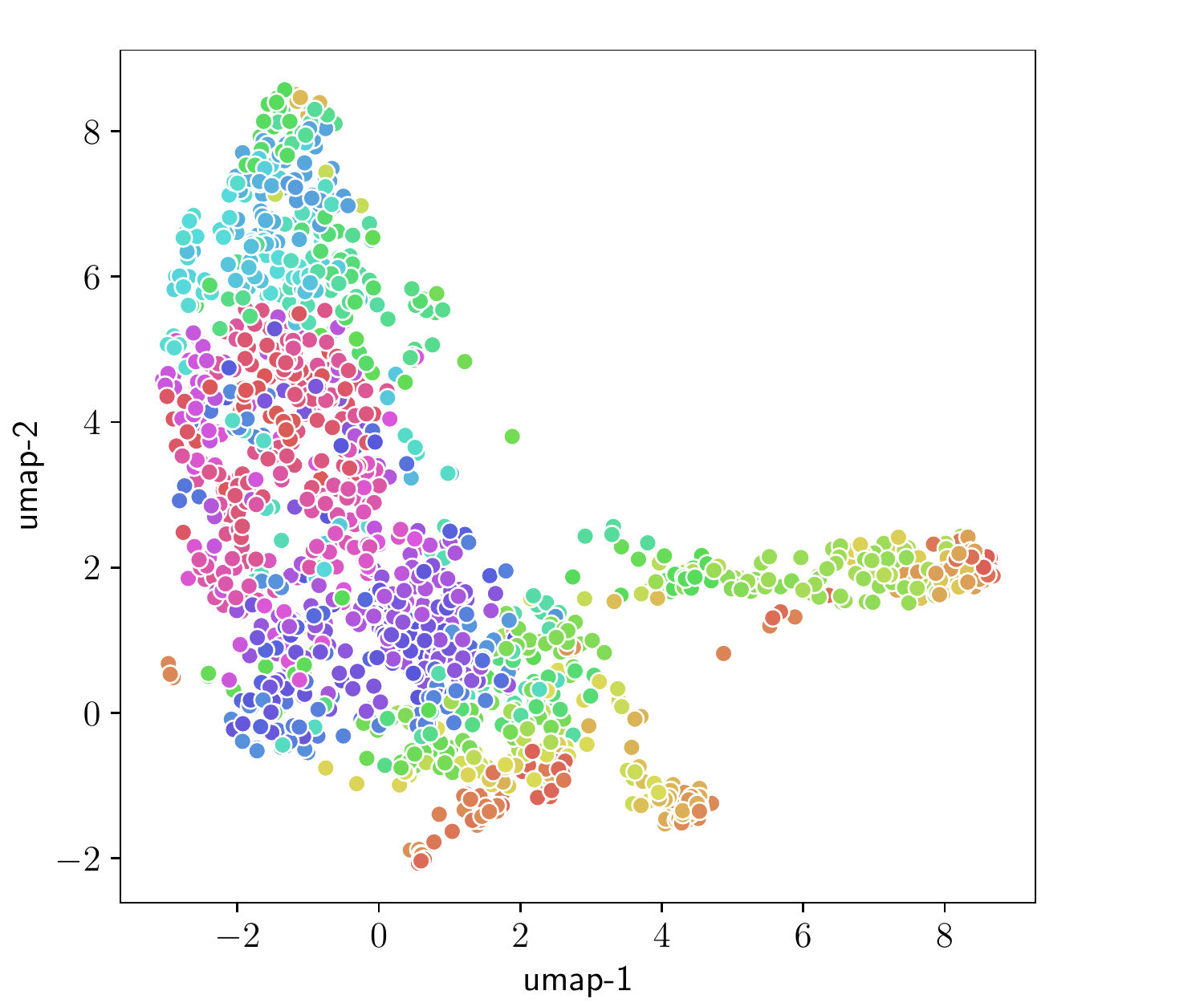}
\caption{UMAP projections of the APOGEE DR16 RC sample coloured by metallicity (left) and depending on the group found by HDBSCAN (right). Member stars of the known OCs are highlighted with different colours and symbols in the left panel}\label{fig:UMAP_APOGEE}
\end{figure*}

We show the two projections of UMAP for all the stars in our sample in Fig.~\ref{fig:UMAP_APOGEE}. The maps present clear patterns, some of which correspond to substructures such as the thick disc (rightmost blob). Very similar patterns are seen in t-SNE representations of the solar vicinity chemical abundance space \citep{Anders+2018}.
The location of the members from the four highlighted clusters indicates that most of their red clump stars have similar abundances, as expected. However, the first impression is that the clusters do not seem clumped enough to be identified as a clear group in the abundance space of APOGEE. Remarkably, two of the clusters (NGC 2682 and NGC 6819) appear to have a very similar abundance pattern, which would make a blind identification more difficult. The most distinctive cluster is NGC 6791. It is located in the most metal-rich region.

\subsection{HDBSCAN on the APOGEE DR16 RC sample}\label{sec:HDBSCAN_APOGEE}
Similarly, as for the sample of high-precision clusters (Sect.~\ref{sec:data}), we now ran HDBSCAN in the APOGEE DR16 RC sample in the chemical space of 18 elements described in Sect.~\ref{sec:APOGEE}.

We performed a similar procedure as in Sect.~\ref{sec:chemtag} (Fig.~\ref{fig:recovery}) to select the HDBSCAN parameters that maximised the recovery of the clusters. The parameters \minsize\ and \minsamp = 3 return the highest recovery fractions, and the choice of \eps\ does not affect the results if (0.01$<$\eps$<$0.1).

In the configuration described above, HDBSCAN finds 272 groups, which are represented in the right plot of Fig.~\ref{fig:UMAP_APOGEE} with the UMAP projections. We find that stars from two different clusters appear in some of the groups that are listed in Table~\ref{tab:HDBS_clusters_APOGEE}. The recovery is very poor. Only the cluster NGC 188 is above the threshold of 40\% on completeness and homogeneity, and no cluster is recovered above the threshold of 70\%.

\begin{table}
\caption{Groups found by HDBSCAN that contain any star from the original clusters in the APOGEE DR16 RC sample. The columns are the same as in Table~\ref{tab:HDBS_clusters}.}\label{tab:HDBS_clusters_APOGEE}
\centering
\setlength\tabcolsep{2pt}
\begin{tabular}{llrrr}
 \hline
HDBSCAN group & Real Cluster(s) & Comp. & Hom.  \\
($N_{\mathrm{FOUND}}$)  & ($\nicefrac{N_{\mathrm{FOUND}}}{N_{\mathrm{REAL}}}$) & & \\
 \hline
H230 (3)        & \textbf{NGC 188}  (2/4)  & 50\% & 66\% \\
H240 (11)       & NGC 6819 (3/11) & 27\% & 27\% \\
H242 (6)        & NGC 6819 (1/6)  & 9 \% & 16\% \\
\hline
\end{tabular}
\end{table}

\subsection{Restricting the chemical space}

For the purpose of chemical tagging, it is not clear whether using several chemical elements that are representative of the same nucleosynthetic path improves or worsens the results.

For instance, \citet{PriceJones+2020} chose to restrict the APOGEE chemical space to eight abundance ratios, [Mg/Fe], [Al/Fe], [Si/Fe], [K/Fe], [Ti/Fe], [Mn/Fe], [Ni/Fe], and [Fe/H], for their chemical tagging. This selection was made based on a simulation that tested the median homogeneity of the recovered groups as a function of the dimensions of the chemical space. With this selection, they included elements coming from mainly SN type II (Mg), others with also a partial contribution from SN type I (Si and Ti), odd-Z elements (K and Al), and iron-peak elements primarily produced by SN type I with an additional contribution from SN type II (Mn, Ni, and Fe). However, in their simulations, they showed that very similar homogeneity scores are also obtained for higher dimensions of the chemical space, up to 15 dimensions.

Recently, \citet{Ting+2021} concluded that at least seven elements have to be considered in APOGEE to remove residual correlations with the purpose of Galactic archaeology studies. They proposed that using Fe, Mg, O, Si, Ca, Ni, and Al might allow them to explain the diversity of abundance patterns of their data, which are composed of disc stars with approximately solar metallicity. The proposed elements are also those with the best measurement uncertainties in APOGEE. The authors mentioned that this is a lower conservative limit in the possible elements to be used and that an analysis including elements produced by other processes will exhibit a richer structure in the data.

We wish to test whether restricting the number of elements to the most significant ones can have a positive effect on cluster recovery. This might be important if the inclusion of certain elements introduces noise in the chemical space due to the uncertainties underlying the computation of abundances (NLTE, faint lines, and poor atomic characterisation).

We ran HDBSCAN in the same configuration as in Sect.~\ref{sec:HDBSCAN_APOGEE} using the chemical space proposed by \citet{PriceJones+2020}, and that proposed by \citet{Ting+2021}. The algorithm finds 365 and 483 groups, respectively. The results of the groups that contain any star from a real cluster are detailed in Table~\ref{tab:HDBS_clusters_APOGEEsel}. The results for both sets of abundances do not seem to improve with respect to the previous test, shown in Table~\ref{tab:HDBS_clusters_APOGEE}. The only cluster that seemed moderately recovered in the previous test using all abundances in terms of our indicators (NGC 188) now is not recovered. For cluster NGC 6819, for which three stars were grouped before, now only two stars are grouped. Apparently, the larger the chemical space, the better the performance. However, it is difficult to really judge the performance of the three sets of abundances given the poor recovery fractions of the results in any case. Moreover, the recovery of one or two stars of the input clusters from more than 300 groups might be attributed to chance rather than to a real spotting or identification of the clusters.

\begin{table}
\caption{Same as Table~\ref{tab:HDBS_clusters_APOGEE}, but with the results of the HDBSCAN run on the set of abundances of \citet{PriceJones+2020} (top) and those of \citet{Ting+2021}.}\label{tab:HDBS_clusters_APOGEEsel}
\centering
\setlength\tabcolsep{2pt}
\begin{tabular}{llrrr}
 \hline
HDBSCAN group & Real Cluster(s) & Comp. & Hom.  \\
($N_{\mathrm{FOUND}}$)  & ($\nicefrac{N_{\mathrm{FOUND}}}{N_{\mathrm{REAL}}}$) & & \\
 \hline
H284 (9)        & NGC 188  (1/4)   & 25\% & 11\% \\
H293 (4)        & NGC 6819 (1/11)  &  9\% & 25\% \\
H328 (5)        & NGC 2682 (1/5)   & 20\% & 20\% \\
H330 (6)        & NGC 2682 (1/5)   & 20\% & 33\% \\
 \hline
H405 (5)        & NGC 6819 (2/11)  & 18\% & 40\% \\
H411 (6)        & NGC 6819 (1/11)  &  9\% & 16\% \\
H430 (3)        & NGC 2682 (1/5)   & 20\% & 33\% \\
H463 (9)        & NGC 188  (1/4)   & 25\% & 25\% \\
 \hline
\end{tabular}
\end{table}

\section{Discussion}\label{sec:discussion}
We have tested the possibilities of chemical tagging using known OCs in two distinct scenarios. The first scenario was  the ideal case of high-precision differential chemical abundances of 175 highly probable member stars from 31 clusters (Sect.~\ref{sec:EVOC}). We tested the recovery fractions of HDBSCAN on this sample with 16 chemical species and also restricted this to the subsample of clusters with more member stars.
As a second test case, we studied the possibility of recovering known clusters in the APOGEE DR16 RC sample (Sect.~\ref{sec:APOGEE}). In this case, we tested the effect of using different chemical spaces, according to recent results by \citet{PriceJones+2019} and \citet{Ting+2021}. Our main findings from these experiments can be summarised as follows:

\begin{enumerate}
 \item We have shown that in the high-precision sample (considered as the best-case scenario, Sect.~\ref{sec:EVOC}), we can recover only 9 out of the 31 analysed clusters, as shown in Table~\ref{tab:HDBS_clusters} and Fig.~\ref{fig:UMAP}. This is obtained when a relaxed threshold of 40\% in the recovered completeness and homogeneity was set, meaning that some of the \emph{correctly} recovered clusters have fewer than half of the real number of cluster stars. The other 22 groups do not represent real clusters and contain a mixture of stars belonging to different clusters. 
With a more restrictive threshold of 70\% \citep{PriceJones+2019}, only one cluster is recovered.

 \item The possibility of recovering known clusters in the chemical space of 18 elements in APOGEE is a less favourable scenario with larger uncertainties. However, it lets us test the performance of the clustering in the presence of field stars. Out of the four clusters with more members in APOGEE, only one is recovered at a threshold of 40\%. We investigated how this performance might change when fewer elements were used,  but the representation of most nucleosynthetic paths was kept. In this case, the recovery was even poorer. No cluster was correctly recovered.
\end{enumerate}

These results point to a difficult interpretation of the eventual groups obtained from a blind clustering search of a large sample of field stars. According to our tests in (1), we can expect that 70\% of the detected groups are statistical overdensities that have arisen from the confusion of the cluster signatures in the chemical space. For case (2), the statistics are even lower, with only one out of 272 groups being a real known cluster.

As a drawback of our experiment, only few of the known OCs in both cases contain a large number of member stars: for instance, there are only five clusters with seven or more members in scenario (1). In Sect.~\ref{sec:restr_sample}, we therefore repeated the same experiment restricted to the most populated clusters (five clusters and 44 stars), which allowed us to test whether it might be easier to find clusters when they have a large number of stars. In this case, we obtained a recovery fraction of 20\% (at a threshold of 70\% for the homogeneity and completeness), which we consider still very poor in this unrealistic and simplistic case.

Our results contrast with the findings by \citet{PriceJones+2019}. It is true, however, that all the candidate birth clusters found by \citet{PriceJones+2019} have at least 15 members. This is probably because there was no selection of the RC in their search. In our test case, all clusters have fewer members because there are very few OCs with more than $\text{about seven}$ members in the RC. Only the oldest and most massive clusters have more than this number of members. The cross-match between the 360 candidate cluster members by \citet{PriceJones+2019} with our APOGEE DR16 RC sample gives 17 stars. Only 5 of these stars appear in the groups found in Sect.~\ref{sec:APOGEE}, and each is assigned to a different group.

A possible point of view might be that if this algorithm were applied to a large sample of stars, it might be easier to determine real dissolved clusters if they were very massive, which would mean that more stars would represent them. The drawback of this idea is, however, that generally, stars in the same evolutionary state are required to perform a meaningful clustering. The retrieved chemical abundances might otherwise have biases among different stars, in addition to the uncertainties due to limited precision. 
In real life, this is only feasible by selecting the RC stars according to a compromise of them being bright (so that larger distances can be reached and a larger sample of stars can be obtained) and providing among the highest precision abundances (needed for strong chemical tagging). A similar precision is also retrieved for GK dwarfs, but these are in general faint and thus limit the sample.
Additionally, only old and massive clusters have a prominent RC population. However, it has been shown that the more massive the cluster, the lower the chance to be dissolved \citep[e.g.][]{Lamers+2005}, but the survival probability also depends on the orbit \citep[see][]{MartinezMedina+2018}.
By a blind strong chemical tagging search, we can therefore probably only hope to find either dissolved clusters that have some chemical peculiarity, or the very few old and massive clusters that are expected to be dissolved in the field. This is shown in our first experiment in Sect.~\ref{sec:EVOC}, where we successfully recovered the clusters NGC~6705 (moderately metal rich and alpha enhanced) and NGC~2420 (the most metal-poor cluster) with the highest rates, even with different configurations of the HDBSCAN free parameters. However, we might be able to find more clusters using kinematical information in addition to chemistry. This is still to be tested and probably would only be possible with the youngest clusters, where the dissipative processes of the Galatic disc have still not fully erased the kinematical similarity of individual stars.

Nevertheless, we add a word of caution to strong chemical tagging: when clusters in a sample are searched for, it is highly probable that clusters are found. However, it is very difficult to understand whether they were once gravitationally bound. Our experiment tells us that we can expect that at least 70\% of the groups are not real birth clusters. This percentage will possibly have a high dependence on the clustering method and abundance precisions involved. In any case, our results do not prevent the use of clumpiness of the chemical space to understand underlying processes of star formation and chemical evolution \citep{Ting+2016}.

In conclusion, we are not able to fully identify which stars belong to which cluster using chemistry alone, even in the best-case scenario of high-quality chemical abundances of 16 chemical species. This result challenges the prospects of strong chemical tagging, at least using the current abundance precision obtained from the available wavelength range at high resolution.

\section{Conclusions}\label{sec:conclusions}

Well-known clusters provide the best test case to investigate whether strong chemical tagging is possible. Being able to chemically tag groups in a large sample of field stars would provide a way to allow temporal sequencing of a large fraction of stars in our Galaxy in a manner analogous to building a family tree. The promise of chemical tagging has been one of the strongest arguments used to justify current and future spectroscopic surveys. It is therefore important to study it with controlled samples. Two assumptions are needed for it to work: the members of a birth cluster should have a chemically homogeneous composition, and each cluster should have a unique chemical signature to be able to distinguish stars from different clusters. We find strong evidence against the second hypothesis.

We investigated the feasibility of strong chemical tagging using two samples of known clusters to test two possible scenarios. 

In the first case, we used chemical abundances of 175 stars in 31 clusters obtained by \citet{Casamiquela+2021}, with ages ranging from 200 Myr to 7 Gyr. The chemical space consisted of elements coming from different nucleosynthetic paths: Na, Al, Mg, Si, Ca, Sc, Ti, V, Cr, Mn, Co, Fe, Ni, Zn, Y, and Ba. All stars corresponded to red clump stars, and individual stellar abundances are smaller than 0.05 dex. No field stars were present in this sample.
Even when the internal coherence of the stellar abundances in the same cluster was high, typically 0.03 dex, we observed that the overlap in the mean chemical signatures of the clusters is large.

We applied the clustering algorithm HDBSCAN in our sample. We fine-tuned the free parameters of the algorithm to obtain the best results in terms of completeness, homogeneity, and recovery fraction of the groups found. Our best results have an overall completeness, homogeneity, and V-measure of 63\%, 49\%, and 55\%, respectively. In terms of how the individual clusters were recovered, we considered 29\% of the sample recovered (nine clusters) at a 40\% threshold on completeness and homogeneity, and only one cluster was recovered (NGC~2682, 3\% of the sample) at a threshold of 70\%.
The UMAP representation of the chemical space clearly shows that there is a large mixture of stars from similar clusters in the central parts of the distribution. The clusters recovered in HDBSCAN groups are preferentially found at the edges of the distribution in the UMAP space.

In the second case, we used the APOGEE DR16 RC sample with the abundances computed by \emph{astroNN}. We tried to recover the known clusters embedded in this catalogue using the same algorithm, HDBSCAN. Using the chemical space of 18 elements, the algorithm found 272 groups. Only one is considered recovered at a threshold of 40\% homogeneity and completeness. In this case, we also tested whether the restriction of the chemical space can help in the identification of real clusters. We find that there is no improvement in the results. 

Overall, our results show that the chances of recovering a large fraction of clusters dissolved in the field are slim because in our best-case scenario without field stars, more than 70\% of the groups of stars are in fact statistical groups that contain stars belonging to different real clusters. This is probably because the overlap in the chemical signatures of OCs is large, and the chemistry of the thin disc has a very small range in abundance compared with the precision. We showed, however, that some clusters are persistently recovered. They have the particularity of being at the edges of the distribution in the UMAP projections (e.g. NGC~2420, NGC~6705, and NGC~2682). Thus, we can hope to recover some of the birth clusters that have a particular chemistry that stands out of the general distribution of the thin disc.

This shows how challenging it is to apply a blind clustering search to the chemical space of a large group of field stars from a large spectroscopic survey. We conclude that it will be difficult to interpret if the recovered groups come from real birth clusters, or if in fact they are statistical overdensities.

\begin{acknowledgements}
We thank Laura Magrini for refereeing this paper and for her suggestions that improved the work.

This study has made use of data from the European Space Agency (ESA) mission \emph{Gaia} (\url{http://www.cosmos.esa.int/gaia}), processed by the \emph{Gaia} Data Processing and Analysis Consortium (DPAC, \url{http://www.cosmos.esa.int/web/gaia/dpac/consortium}). We acknowledge the \emph{Gaia} Project Scientist Support Team and the \emph{Gaia} DPAC. Funding for the DPAC has been provided by national institutions, in particular, the institutions participating in the \emph{Gaia} Multilateral Agreement.
This research made extensive use of the SIMBAD database, and the VizieR catalogue access tool operated at the CDS, Strasbourg, France, and of NASA Astrophysics Data System Bibliographic Services.

We acknowledge support from "programme national de physique stellaire" (PNPS) and the "programme national cosmologie et galaxies" (PNCG) of CNRS/INSU. L.C. acknowledges the support of the postdoc fellowship from the French Centre National d’Etudes Spatiales (CNES). FA acknowledges funding from MICINN (Spain) through the Juan de la Cierva-Incorporación program under contract IJC2019-04862-I.

We also thank undergraduate students Jaume Dolcet and Ignacio García-Soriano for collaborating in preliminary studies for this work. 

\end{acknowledgements}

\bibliographystyle{aa} 
\bibliography{biblio2_v2} 

\end{document}